\documentclass[notitlepage]{revtex4-1}

\usepackage{graphicx}
\usepackage{enumerate}
\usepackage{amsmath}
\usepackage{amssymb}
\usepackage{fancyhdr}
\usepackage[hidelinks=true]{hyperref}  
\usepackage{braket}
\hypersetup{
    colorlinks,
    citecolor=red,
    filecolor=black,
    linkcolor=blue,
    urlcolor=black
}

\begin{document}
\title{Quantum Dynamics of Complex Hamiltonians}
\author{Kushagra Nigam}
\email{f2010174@goa.bits-pilani.ac.in}
\affiliation{Department of Physics, BITS Pilani K.K Birla Goa Campus, N.H. 17B Zuarinagar, Goa 403726, India.}
\author{Kinjal Banerjee}
\email{kinjalb@gmail.com}
\affiliation{Department of Physics, BITS Pilani K.K Birla Goa Campus, N.H. 17B Zuarinagar, Goa 403726, India.}

\date{\today}

\begin{abstract}
\textbf{Abstract:} 
Non hermitian Hamiltonians play an important role in the study of dissipative quantum systems. We show that using states
with time dependent normalization can simplify the description of such systems especially in the context of the
classical limit. We apply this prescription to study the damped harmonic oscillator system. This is then used to study
the problem of radiation in leaky cavity. 
\end{abstract}

\maketitle

\vspace{-5ex}
\tolerance=1 
\emergencystretch=\maxdimen
\hyphenpenalty=10000
\hbadness=10000
\section{Introduction}

Quantum Mechanics in its standard form deals with systems like free particle, simple harmonic oscillator, Hydrogen atom etc., all
of which exhibit conservation of energy. Even when we want to account for interaction with external fields like magnetic field 
or electric field the number of degrees of freedom that facilitate energy exchange is small and hence can be formulated using a 
Hamiltonian with real eigen values. However, when it comes to the case of dissipative systems, the degrees of freedom responsible 
for system-environment interaction grow enormously \cite{feynman} and standard tools of quantum mechanics are not sufficient. 
This makes the problem of having a quantum mechanical description of dissipative systems extremely challenging and
difficult \cite{yan}. This has been an area of active study and several tools have been developed to describe various
aspects of quantum systems in the presence of dissipative effects (see \cite{dissipationreviews} and references therein). 

The damped harmonic oscillator provides a toy model to study dissipative systems. There have been several attempts to
obtain a consistent quantum theory with the correct classical limit of this system. Some of the approaches which have
been notably successful include Bateman's dual oscillator approach and the Caldirola-Kanai Hamiltonian with time
dependent mass. A review of the approaches with a comprehensive list of references can be found in \cite{dekker}. More
recently, this system has been analyzed from quantum field theory perspective in \cite{QFTDHO} and via Feynman path
integral in \cite{PIDHO}. Some of the recent studies of the damped oscillator have been carried out in \cite{recentdho}
(also see \cite{dhoreview}). Extensive work has been done to model dissipation by coupling system 
and environment using a system-environment coupling term in the Hamiltonian \cite{senitzky}.
  
Since dissipation involves energy being transferred out of the system into the environment, there have been efforts to
try to describe such systems using Hamiltonians which are non-hermitian. Traditionally non-hermitian Hamiltonians have 
been used to phenomenologically describe processes which lead to a non conservation on particle number, eg radioactive
decay, scattering  and transport phenomena \cite{nherhamrev1}. Although, in most cases, non-Hermitian Hamiltonians are 
phenomenological which are introduced heuristically, they give a mathematical consistent formalism of describing a
system coupled with a continuum of states \cite{nherhamrev2}. Understanding such Hamiltonians and their
properties and applications is a current area of active research from both theoretical and experimental perspectives 
\cite{nhhth,nhhp}. In particular, they have been used to study damped oscillator systems \cite{Graefe}.  

It is important to stress that if we consider the system of interest and the environment together as a single
system, then that system is not dissipative. However such a description, though complete, will not be of much use 
unless we manage to \textit{integrate out} the environmental degrees of freedom. That is not the path we
take in this paper. Here we try to build up tools which can provide an \textit{effective} quantum description of
only the dissipative system. We will provide a prescription for constructing quantum mechanical description of
dissipative systems using non-hermitian Hamiltonians similar to the path followed in \cite{Graefe} although our
motivation and approach is quite different. Our goal is the following: 
given a classical (dissipative) system, we propose a quantum (non-hermitian) Hamiltonian
operator such that, for a given class of states, the expectation values of certain operators are peaked about their
classical values. The operators we are concerned with are the ones corresponding to classical measurables. We use
coherent states of the corresponding non-dissipative systems to calculate the expectation values. The justification for
the correctness of this proposed construction is that the expectation values agrees with the ones calculated using
other methods in literature. In addition to the simplicity of the procedure, another major advantage of our construction
is that in the limit where the dissipation goes to zero, our results naturally goes over to the results of the
conservative system. 

In this paper, we will also deal with the leaky cavity which is another dissipative system that finds its application in
some areas of experimental physics including laser physics and quantum optics \cite{lcref}. Usually the quantity of interest in dealing
with such systems is the radiation loss due to leakage. A lot of progress has been made in the study of quantization
modes of leaky cavities (see \cite{dutra1} and references therein) and it is not possible to summarize it in one
paragraph. Very briefly, one of the more popular approaches is using the 
Gardiner-Collett Hamiltonian \cite{GCH} where the cavity field is described in terms of normal modes as if the
cavity were perfect and closed. The leakage is modelled by introducing an ad-hoc coupling with a reservoir. Finally the
reservoir states are integrated out to obtain a master equation which contains information about the modes in the
cavity. We are not really interested in the technical aspects of this problem but in this paper we show how our formalism
for the damped oscillator can be directly mapped onto the problem of the leaky cavity which can then be solved in the
same way.
  
 This paper is organized as follows:
In section (\ref{sec:Dissipat}) we start by reviewing a general quantum mechanical formalism for a complex Hamiltonian
systems. Section (\ref{sec:DHO}) illustrates the application of formalism so developed to model damped harmonic oscillator (DHO). 
Our proposal, makes an improvement to the scheme suggested in \cite{Graefe} which greatly simplifies the procedure of quantum-classical 
correspondence using Ehrenfest's theorem. In section (\ref{sec:leaky}) section we investigate the phenomenon of radiation leakage in cavity 
using complex Hamiltonians. We postulate Hamiltonian that allows for polarization dependent decay factors and then derive the time evolution
of Electric and Magnetic fields. This gives a novel treatment of the leaky cavity problem which has not appeared in
literature before. We finally end with conclusions in section (\ref{sec:conclude}).


\section{Quantum Theory of Dissipative Systems}\label{sec:Dissipat}

In this section, we explore some of the features of a general non-hermitian Hamiltonian. Since most dissipative systems
are modeled by adding a dissipative part to an initially conservative system, we will also assume that our Hamiltonian has 
distinct conservative and dissipative pieces. We will be closely following the treatment given in \cite{Graefe}. 
Let us start with the following Hamiltonian
\begin{equation}\label{dissiphamilt}
\hat{H}_{D}=\hat{H} - i\hat{\Gamma}
\end{equation}
where, in anticipation we have added the subscript $D$ to stand for dissipative system. We further require the following conditions to be satisfied
\begin{itemize}
\item $\hat{H}=\hat{H}^\dagger$ where, $\hat{H}$ is Hamiltonian of undamped system 
\item $\hat{\Gamma}=\hat{\Gamma}^\dagger$ where, $\hat{\Gamma}$ is damping part of Hamiltonian 
\item $[\hat{H},\hat{\Gamma}]=0$ 
\end{itemize}
The third condition is to ensure the existence of simultaneous eigenkets for $\hat{H}$ and $\hat{\Gamma}$. This will
also ensure that the Hamiltonian  (\ref{dissiphamilt}) will have a complex eigenvalues. 

Now, given a quantum system described by state $\ket{\psi}$ its time evolution as governed by the time dependent Schroedinger equation
\begin{align}\label{diftimevol}
i\hbar\frac{\partial \ket\psi}{\partial t}&=\hat{H}_D\ket{\psi} \\
\implies \ket{\psi,t+dt}&=\left(1-i\frac{\hat{H}_{D}dt}{\hbar}\right)\ket{\psi,t}=\hat{U}(t,t+dt)\ket{\psi,t}
\end{align}
where, $\hat{U}(t,t+dt)$ is now a non-unitary time evolution operator.

Since, Hamiltonian $\hat{H}$ is time independent this implies that Hamiltonians at different time commute. 
Thus, we can find out finite time evolution of state $\ket{\psi}$ for time $t$ as
\begin{align}\label{fintimevol} 
\ket{\psi,t}&=\lim\limits_{N\rightarrow\infty}\left(1-i\frac{\hat{H}_{D}t}{\hbar N}\right)^N\ket{\psi,t}=
e^{-i\frac{\hat{H}_{D}t}{\hbar}}\ket{\psi,0}=\hat{U}(t)\ket{\psi,0}
\end{align}
It is well known that probability is not conserved under evolution by a non unitary operator. To see that consider 
an initially normalized state $\ket{\psi}$
\begin{align}\label{state}
\ket{\psi,t=0} = \sum\limits_{n=0}^\infty c_n\ket{n} ~ ~ ~ ;  ~ ~ \sum\limits_{n=0}^\infty |c_n|^2=1
\end{align}
where we have expanded the state $\ket{\psi}$ in terms of the energy eigenstates of the undamped system $\ket{n}$. Note
that we are assuming that the energy eigenstates of the undamped system span the whole Hilbert space. Also note that
since $[\hat{H},\hat{\Gamma}]=0$, the kets $\ket{n}$ are simultaneous eigenkets of $\hat{H}$ and $\hat{\Gamma}$.

Under time evolution (\ref{fintimevol}) we have
\begin{align}\label{evol}
\ket{\psi,t}&=\hat{U}(t)\ket{\psi,0}=e^{-i\frac{\hat{H}_{D}t}{\hbar}}\ket{\psi,0} \\
&=\sum\limits_{n=0}^\infty c_ne^{-\frac{\Gamma_n t}{\hbar}}e^{-i\frac{{E}_{n}t}{\hbar}}\ket{n}
\end{align}
where $E_n$ and $\Gamma_n$ are the eigenvalues of  $\hat{H}$ and $\hat{\Gamma}$ respectively.

Since, time evolution is non unitary in nature, the norm is no longer time independent and must evolve according to
\begin{align}\label{tdepnorm}
N(t)=\braket{\psi,t|\psi,t}=\sum\limits_{n=0}^\infty |c_n|^2e^{-2\frac{\Gamma_n t}{\hbar} }
\end{align}
The time derivative of the norm is given by
\begin{align}\label{derivnorm}
\hbar\frac{dN(t)}{dt}=-2\braket{\psi,t|\hat{\Gamma}|\psi,t}
\end{align}
To obtain a consistent physical interpretation in such cases we will have to slightly modify the notion of time evolution of expectation values of
any operator. Given any observable $\hat{A}$, we define its expectation value at time $t$ with respect to state
$\ket{\psi,t}$ to be
\begin{align}\label{expectop}
\braket{\hat{A}}_t = \frac{\braket{\psi,t|\hat{A}|\psi,t}}{N(t)} := \braket{\phi,t|\hat{A}|\phi,t} 
\end{align}
where we have defined the state, $\ket{\phi,t}:=\frac{\ket{\psi,t}}{\sqrt{N(t)}}$ which is normalized for all times.
 
To find (\ref{expectop}) we can use (\ref{fintimevol}) to obtain
\begin{eqnarray}\label{expevol}
&&\bra{\psi,t+dt}\hat{A}\ket{\psi,t+dt} = \bra{\psi,t}\hat{U}^\dagger(dt)\hat{A} ~ \hat{U}(dt)\ket{\psi,t} \nonumber \\
&\Rightarrow& i\hbar\frac{d\bra{\psi,t}\hat{A}\ket{\psi,t}}{dt} =
\bra{\psi,t}[\hat{A},\hat{H}]\ket{\psi,t}-i\bra{\psi,t}\left\{\hat{\Gamma},\hat{A}\right\}\ket{\psi,t} 
\end{eqnarray}
where, $[\hat{A},\hat{H}]:= \hat{A}\hat{H}-\hat{H}\hat{A}$ and
$\left\{\hat{\Gamma},\hat{A}\right\} := \hat{\Gamma}\hat{A}+\hat{A}\hat{\Gamma}$.

Using (\ref{derivnorm}) and (\ref{expevol}) we find the time derivative of expectation value of operator $\hat{A}$ as
\begin{align}\label{expectopevol}
 i\hbar\frac{d\bra{\phi,t}\hat{A}\ket{\phi,t}}{dt}&=\bra{\phi,t}[\hat{A},\hat{H}]\ket{\phi,t}-2i\Delta^2_{A\Gamma}
\end{align}
where,
$\Delta^2_{A\Gamma} := \bra{\phi,t}\frac{1}{2}\left\{\hat{A},\hat{\Gamma}\right\}\ket{\phi,t}
-\bra{\phi,t}\hat{A}\ket{\phi,t}\bra{\phi,t}\hat{\Gamma}\ket{\phi,t}$.

Note that, as expected, all the deviations from standard quantum mechanics involve only the non-hermitian term
$\hat{\Gamma}$ and its eigenvalues. When the dissipation goes to zero we get back our standard quantum mechanics.

We further observe that the Schroedinger equation (\ref{diftimevol}) can be modified to account for time dependent normalization factor. 
The modified non-linear Schroedinger equation can be written in terms of normalized states $\ket{\phi}$ using (\ref{derivnorm}) as 
\begin{align} \label{modifschro}
i\hbar\frac{d\ket{\phi,t}}{dt}=\hat{H}\ket{\phi,t} - i[\hat{\Gamma},\ket{\phi,t}\bra{\phi,t}]\ket{\phi,t}
\end{align}
where, $[\hat{\Gamma},\ket{\phi,t}\bra{\phi,t}]=\hat{\Gamma}\ket{\phi,t}\bra{\phi,t}-\ket{\phi,t}\bra{\phi,t}\hat{\Gamma}$ 
and $\ket{\phi,t}\bra{\phi,t}$ is the projection operator on state $\ket{\phi,t}$.\newline 

This is a brief overview of the modifications we obtain in the standard quantum mechanics, if we start with an explicit
non-hermitian Hamiltonian. So far, it is just a mathematical curiosity and we have not made any comments on the physical
interpretation. In the subsequent sections we will use this formalism to explore some dissipative
systems. In each of those contexts, the motivation and the interpretation of the additional $\Gamma$ term will become
clear.
 

\section{Damped Harmonic Oscillator}\label{sec:DHO}

The simplest dissipative system in classical theory is Damped Harmonic Oscillator (DHO) whose classical equation of
motion is given by 
\begin{equation}\label{classdho}
\ddot{q} + 2\gamma \dot{q} + \omega^2q = 0 
\end{equation}
where $\gamma$ is damping coefficient while $\omega$ is the natural frequency. In our notation $\dot{q}$ indicates
derivative with respect to time $t$. We will be dealing only with the under-damped case. 
The solution for the above equation of motion is given by
\begin{equation}\label{classsoln}
q(t)=q_oe^{-\gamma t}\cos{\omega_d t} + \left(\frac{\dot{q_o}}{\omega_d} + 
\frac{\gamma q_o}{\omega_d}\right)e^{-\gamma t}\sin{\omega_d t}
\end{equation}
where,
$\omega_d=\sqrt{\omega^2-\gamma^2}$.

There have been multiple efforts to quantize this classical system (see \cite{dekker} and references therein), each with
its own strengths and weaknesses. In this section we will propose a quantization scheme, based on the formalism developed
in the previous section, for the DHO. We will postulate a non-hermitian Hamiltonian for the DHO and show that, for a certain 
class of states, we recover the classical equations of motion via Ehrenfest theorem. Another attractive feature of our quantum theory is
that it naturally reduces to theory of Simple Harmonic Oscillator(SHO) as damping coefficient $\gamma\rightarrow0$.

We propose the following Hamiltonian operator for the DHO
\begin{align} \label{dhohamilt}
\hat{H}_D=&\left(\hat{a}^\dagger\hat{a} + \frac{1}{2}\right)\hbar\omega_d - i\hbar\gamma\hat{a}^\dagger\hat{a} \nonumber\\
=&\hat{H}-i\hbar\gamma\hat{\mathcal{N}}
\end{align}
where $\hat{\mathcal{N}}$ is the number operator and $\gamma$ has dimensions of $[T]^{-1}$. Comparing with formalism of section
(\ref{sec:Dissipat}) we can see that the dissipative part of Hamiltonian is taken to be
$\hat{\Gamma} := \hbar\gamma\hat{\mathcal{N}}$. 
 
In an attempt to attain correct classical limit we shall use the coherent states of SHO (\ref{coherent}). The time evolution 
of a coherent state can be easily calculated using (\ref{fintimevol})
\begin{align}\label{cohevol}
\ket{\alpha,t}=\hat{U}(t)\ket{\alpha,0}=e^{-i\frac{\hat{H}_{D}t}{\hbar}}\ket{\alpha,0}
=e^{-\frac{|\alpha^2|}{2}}\sum\limits_{n=0}^\infty \frac{\alpha^n}{\sqrt{n!}}e^{-\gamma nt}e^{-i\frac{{E}_{n}t}{\hbar}}\ket{n}
\end{align}
Obviously the norm will not be preserved under time evolution. The time dependent norm will now be given by (\ref{tdepnorm}) 
\begin{align}\label{norm}
N(t)=\braket{\alpha,t|\alpha,t}=e^{-|\alpha|^2}e^{\left({|\alpha|^2e^{-2\gamma t}}\right)}
\end{align}
The states we are interested in are the normalized coherent state which we label as $\tilde\alpha$, are given
by
\begin{align}\label{normalcoh}
\ket{\tilde{\alpha},t}=
e^{-\left(\frac{|\alpha|^2}{2}e^{-2\gamma t}\right)}\sum\limits_{n=0}^\infty \frac{\alpha^n}{\sqrt{n!}}e^{-\gamma nt}e^{-i\frac{{E}_{n}t}{\hbar}}\ket{n}
\end{align}
These are the states which are physically relevant.  All the 
expectation values calculated subsequently in this section are with respect to these states. 
In particular, the time dependent normalization naturally ensures that as time progresses the probability of finding 
the system in higher excited state decreases and the probability of finding the 
oscillator in lower excited states simultaneously increases. This is exactly what we hope to model while studying a
dissipative system. 

It is well known that coherent states are minimum uncertainty states for a simple harmonic oscillator. We are using the
same coherent states (with an additional time dependent normalization) to study the damped oscillator. To understand the
nature of these states for our system let us calculate the expectation value of position and momentum operator 
\begin{align}\label{xpmean}
q(t)=\bra{\tilde{\alpha},t}\hat{q}\ket{\tilde{\alpha},t}&=
\bra{\tilde{\alpha},t}\sqrt{\frac{\hbar}{2m\omega_d}}(\hat{a}+\hat{a}^\dagger)\ket{\tilde{\alpha},t} 
=\sqrt{\frac{2\hbar}{m\omega_d}}\;\;\mbox{Re}\left(\alpha e^{-it\omega_d}\right)e^{-\gamma t} \\
p(t)=\bra{\tilde{\alpha},t}\hat{p}\ket{\tilde{\alpha},t}&=
\bra{\tilde{\alpha},t}\sqrt{\frac{m\omega_d\hbar}{2}}\left(\frac{\hat{a}-\hat{a}^\dagger}{i}\right)\ket{\tilde{\alpha},t} 
=\sqrt{2m\omega_d\hbar}\;\;\mbox{Im}\left(\alpha e^{-it\omega_d}\right)e^{-\gamma t}
\end{align}
Since, $\alpha$ in general is a complex number, its real and imaginary part essentially contains 
information about initial position $q_0$ and momentum $p_0$ respectively, i.e.
\begin{align}\label{alphas}
\mbox{Re}(\alpha) = \sqrt{\frac{m\omega_d}{2\hbar}}q_{o} \hspace{2em};\hspace{2em}
\mbox{Im}(\alpha) = \sqrt{\frac{1}{2m\omega_d\hbar}}p_{o}
\end{align}
We further observe that the variances of position and momentum operators are given by
\begin{align}\label{xpvar}
\braket{\Delta\hat{q}^2}&=\frac{\hbar}{2m\omega_d}+\frac{\hbar}{m\omega_d}\left[|\alpha|^2 
+ \mbox{Re}(\alpha^2 e^{-it2\omega_d}) -2\left(\mbox{Re}(\alpha e^{-it\omega_d})\right)^2\right]e^{-2\gamma t} 
=\frac{\hbar}{2m\omega_d} \\
\braket{\Delta\hat{p}^2}&=\frac{m\omega_d\hbar}{2}+{m\omega_d\hbar}\left[|\alpha|^2 
-\mbox{Re}(\alpha^2 e^{-it2\omega_d}) -2\left(\mbox{Im}(\alpha e^{-it\omega_d})\right)^2\right]e^{-2\gamma t} 
=\frac{m\omega_d\hbar}{2}
\end{align}
Therefore, $\ket{\tilde{\alpha},t}$ form minimum uncertainty states.
\begin{equation}
\bra{\tilde{\alpha},t}\Delta\hat{x}^2\ket{\tilde{\alpha},t}\bra{\tilde{\alpha},t}\Delta\hat{p}^2\ket{\tilde{\alpha},t}=\frac{\hbar^2}{4}
\end{equation}
Also note that this uncertainty is independent of time. Thus, the normalized coherent states of SHO (\ref{coherent}) 
with an additional appropriate time dependent normalization form minimum uncertainty states for the DHO. 
These states provide a good description of the damped system for all times.
 
To show that the energy of the damped system actually decays with time let us calculate the expectation value 
$E$ of Hamiltonian corresponding to simple harmonic oscillator with respect to (\ref{normalcoh}).
\begin{align}\label{dhoenergy}
E=\bra{\tilde{\alpha},t}\hat{H}\ket{\tilde{\alpha},t} 
&=e^{- \left(|\alpha|^2e^{-2\gamma t}\right)}\sum\limits_{n=0}^\infty \frac{|\alpha|^{2n}}{n!}e^{-2\gamma tn}\left[n+\frac{1}{2}\right]\hbar\omega_d \nonumber \\
&=|\alpha|^2e^{-2\gamma t}\hbar\omega_d +\frac{1}{2}\hbar\omega_d 
\end{align}
The exponential factor in the first term in equation (\ref{dhoenergy}) goes to $0$ as $t \rightarrow \infty$. 
We can easily calculate the expectation value for the full DHO Hamiltonian (\ref{dhohamilt}) as
\begin{align}\label{dhohamiltexp}
\bra{\tilde{\alpha},t}\hat{H}_D\ket{\tilde{\alpha},t} 
&=\bra{\tilde{\alpha},t}\hat{H}\ket{\tilde{\alpha},t}-i\bra{\tilde{\alpha},t}\hat{\Gamma}\ket{\tilde{\alpha},t}\nonumber\\
&=E+\frac{i\hbar}{2N(t)}\frac{dN(t)}{dt}  ~ ~ ~ ~ ~ ~\mbox{using eq (\ref{derivnorm})} \nonumber\\
&=|\alpha|^2e^{-2\gamma t}\hbar\omega_d +\frac{1}{2}\hbar\omega_d - i\gamma\hbar|\alpha|^2e^{-2\gamma t}
\end{align}
Note that in the second step we have to divide by $N(t)$ as the states considered in equation (\ref{derivnorm}) are not normalized.
The real part of expectation value of $\hat{H}_D$ contains information about decay of energy and the imaginary part
tells about change of norm with respect to time. 

In order to establish connection with classical mechanics using Ehrenfest's theorem let us use equation (\ref{expectopevol}) 
for finding out the evolution of position and momentum operators.
\begin{align}\label{posevol}
 i\hbar\frac{d\braket{\hat{q}}}{dt}&=\braket{[\hat{q},\hat{H}]}-2i\Delta^2_{q\Gamma} \nonumber\\
 \implies  \frac{dq(t)}{dt}&=\frac{p(t)}{m} - \gamma q(t) \\
\label{momevol}
 i\hbar\frac{d\braket{\hat{p}}}{dt}&=\braket{[\hat{p},\hat{H}]}-2i\Delta^2_{p\Gamma} \nonumber\\
 \implies  \frac{dp(t)}{dt}&=-q(t)m\omega^2_d - \gamma p(t)
\end{align}
Differentiating (\ref{posevol}) with respect to time and using (\ref{momevol}) we obtain the following equation
\begin{align}\label{class}
&\ddot{q} + 2\gamma\dot{q} + q(\omega^2_d + \gamma^2) = 0 \nonumber \\
\mbox{or,} ~ ~ ~ ~ &\ddot{q} + 2\gamma \dot{q} + \omega^2q = 0 ~ ~ ~ ~ \mbox{where} ~ ~ \omega^2_d + \gamma^2 = \omega^2
\end{align}
Hence we get back the classical equation of motion which holds for all times. 
It can be easily seen that the formalism reduces to that of SHO in the limit $\gamma\rightarrow0$.

Although the analysis done in this section is similar to the one in \cite{Graefe}, the key difference lies in the choice
of states. We do not use the coherent states $\ket{\alpha}$ for taking the expectation values but the states
$\ket{\tilde{\alpha}}$ which include a time dependent normalization. The advantage of using our states is that the
understanding of classical limit and time evolution of measurable quantities become greatly simplified.
Moreover the expectation values from quantum theory reproduces the correct classical equations of motion which remain
valid for all times.

To recap, in this section we have used the formalism described in the previous section to obtain a quantum description of the damped
oscillator system. The Hamiltonian for the system is non-hermitian (\ref{dhohamilt}). The states we chose to study the
system are simply the coherent states of the simple harmonic oscillator multiplied by a time dependent normalization
factor (\ref{normalcoh}). We show that these states have a very nice interpretation as minimum uncertainty states. If we
view this system as a simple harmonic oscillator which loses energy due to some damping process, the energy 
loss by the dissipative process is demonstrated by the fact that the probability of finding the oscillator in the higher
energy eigenstates decreases with time. We also demonstrate that we obtain the correct classical limit via the
Ehrenfest theorem. One major advantage of this formalism over some of the others \cite{dekker} is that the uncertainty
principle is not violated even for $t \rightarrow \infty$ and the fact that it is trivial to take the limit $\gamma
\rightarrow 0$. This formalism for describing dissipative systems can be applied to more general settings as we shall
show in the next section 
 

\section{Radiation Decay in Leaky Cavitiy}\label{sec:leaky}

The ideal case of radiation inside a cavity problem is treated with cavity being made up of perfectly conducting boundaries \cite{ll}. 
In reality however, cavity walls can have finite transmission coefficient, thus allowing for the radiation to decay or `leak' through.
When considering the leaky cavity, we shall assume that thermal fluctuations can be neglected. we also assume that the decay time is
much longer than the speed of electromagnetic radiation in the cavity. In that case, we can neglect the fact that the
leakage loss is localized at the boundaries and assume that the loss changes the electromagnetic field inside the entire
cavity \cite{dutra1}. Again, this analysis should not be taken as a realistic study of the leaky cavity problem but more
as a demonstration of how non-hermitian Hamiltonians can help tackle such problems in general.

\subsection{Quantization of Radiation}\label{subsec:quantization}
 
In this section we will map leaky cavity problem to that of a damped oscillator and show how it can be used to
obtain how the electromagnetic fields change with time in that case. Our derivation is in the same spirit as done in
\cite{ll} with slight modifications. After obtaining suitable Hamiltonian, we will add a complex part and use the formalism developed 
in previous sections to derive expressions for electric and magnetic fields in presence of leakage. We will then give a physical interpretation of 
decay in terms of photon number. 

Let us consider an ideal  cuboidal cavity. The wave equation that vector potential $\textbf{A}(\textbf{r},t)$ 
has to obey inside the cavity in Radiation gauge is given by
\begin{align}\label{gaugecondA}
\nabla.\textbf{A}&=0  \\
\label{waveq}
\frac{\partial^2 \textbf{A}(\textbf{r},t)}{\partial t^2}&=c^2\nabla^2\textbf{A}(\textbf{r},t)
\end{align}
where, $\nabla^2$ is laplacian operator in 3-Dimensions.
The boundary conditions demand the fields to vanish at boundaries of the cavity. This means that $\textbf{A}(\textbf{r},t)$ 
can admit only standing wave solutions. The most general solution however, can be written as Fourier expansion of standing waves of different wavelengths,
\begin{align}\label{sol1}
\textbf{A}(\textbf{r},t)=\sum_{\textbf{k}=-\infty}^{\infty}\textbf{A}_\textbf{k} \sin(k_x x)\sin(k_y y)\sin(k_z z)\cos(\omega_kt)
\end{align}
where, $\textbf{A}_\textbf{k} (\textbf{r},t)$ are real constant vectors,  $k_x$, $k_y$, $k_z$ are components of wave vector 
$\textbf{k}$ and $\omega_k$ is the angular frequency of wave corresponding to wave vector $\textbf{k}$ and is given as $\omega_k=|\textbf{k}|c$.
If the dimensions of cavity along $x$, $y$, $z$ axis are $l_x$, $l_y$, $l_z$ respectively then boundary conditions demand
\begin{align}\label{boundary}
k_xl_x&=n_x\pi \nonumber\\
k_yl_y&=n_y\pi \nonumber\\
k_zl_z&=n_z\pi
\end{align}
where, $n_x, n_y, n_z \in {\{\mathbb{Z}-\{0\}\}}$ as that there will be no fields corresponding to zero mode due to boundary conditions. 
Expressing the sines and cosines in ($\ref{sol1}$) in terms of exponentials we have
\begin{align}\label{sol2}
\textbf{A}(\textbf{r},t)&=\sum_{\textbf{k}=-\infty}^{\infty}\tilde{\textbf{A}}_\textbf{k} e^{i\textbf{k}.\textbf{r}}\cos(\omega_kt) \\
\label{solstanding}
\implies\textbf{A}(\textbf{r},t)&=\frac{1}{2}\sum_{\textbf{k}=-\infty}^{\infty}\left[\tilde{\textbf{A}}_\textbf{k} 
e^{-i\omega_kt+i\textbf{k}.\textbf{r}}+ \tilde{\textbf{A}}_\textbf{k}e^{i\omega_kt+i\textbf{k}.\textbf{r}}\right] \nonumber\\
&=\sum_{\textbf{k}=-\infty}^{\infty}\left [\tilde{\textbf{b}}_\textbf{k}e^{i\textbf{k}.\textbf{r}}+
\tilde{\textbf{b}}^*_{\bf{-k}}e^{i\textbf{k}.\textbf{r}}\right]
\end{align}
where, $\tilde{\textbf{A}}_\textbf{k}$ are now constant complex 3-dimensional vectors such that 
$\tilde{\textbf{A}}^*_\textbf{k}=\tilde{\textbf{A}}_{(\textbf{-k})}$ and 
$\tilde{\textbf{b}}_\textbf{k}=\frac{1}{2}\tilde{\textbf{A}}_\textbf{k}e^{-i\omega_kt}$. 
Here, we have expressed (\ref{solstanding}) in terms of standing wave solutions for each $\textbf{k}$. This is done so as to create 
mapping between electromagnetic and simple harmonic oscillator system. We further notice that due to (\ref{gaugecondA}), fields have 
two independent direction of polarization for a given direction of propagation $\textbf{k}$.  Let us denote these by 
${\bf{e}}_{\bf {k\alpha}}$ where $\alpha=1,2$. Thus, we can write 
$\tilde{\textbf{b}}_\textbf{k}=\tilde{b}_{\bf{k}_1}{\bf{e}}_{\bf{k}_1} + \tilde{b}_{\bf{k}_2}{\bf{e}}_{\bf{k}_2}$ 
where, $\tilde{b}_{\bf{k}_\alpha}$ are complex scalars. Hence, the standing wave solutions (\ref{solstanding}) can be written as
\begin{align}\label{stand}
\textbf{A}(\textbf{r},t)&=\sum_{\substack{\textbf{k}=-\infty \\ \alpha=1,2}}^{\infty}\left[\tilde{b}_{\bf{k\alpha}}e^{i\textbf{k}.\textbf{r}}
+\tilde{b}^*_{\bf{-k\alpha}}e^{i\textbf{k}.\textbf{r}}\right]{\bf{e}}_{\bf {k\alpha}}
\end{align}
For a electromagnetic system, the energy density is given as 
\begin{align}\label{energy}
\mathcal{H}&=\int\limits_{\mathcal{V}}\left[\frac{1}{2}\epsilon_o|\textbf{E}|^2 +\frac{1}{2\mu_o}|\textbf{B}|^2\right]dV \nonumber\\ 
&=\frac{1}{2}\epsilon_o\int\limits_{\mathcal{V}}\left[|\textbf{E}|^2 +c^2|\textbf{B}|^2\right]dV
\end{align}
where $\mathcal{V}$ is cavity volume.

Since our basic variable is the vector potential $A$ we need to rewrite the $E$ and $B$ fields in terms of $A$ (\ref{solstanding}). 
In the radiation gauge they can be written as
\begin{align}
\label{electric}
{{\bf {E}}({\bf r},t)}&=-\frac{\partial \textbf{A}}{\partial t}=\sum_{\substack{\textbf{k}=-\infty \\ \alpha=1,2}}^{\infty}i\omega_k\left [\tilde{b}_{\bf{k\alpha}}e^{i\textbf{k}.\textbf{r}}-\tilde{b}^*_{\bf{-k\alpha}}e^{i\textbf{k}.\textbf{r}}\right]{\bf{e}}_{\bf {k\alpha}} \\
\label{magnetic} 
{{\bf {B}}({\bf r},t)}&=\nabla \times \textbf{A} =\sum_{\substack{\textbf{k}=-\infty \\ \alpha=1,2}}^{\infty}i\textbf{k}\times\left [\tilde{b}_{\bf{k\alpha}}e^{i\textbf{k}.\textbf{r}}+\tilde{b}^*_{\bf{-k\alpha}}e^{i\textbf{k}.\textbf{r}}\right]{\bf{e}}_{\bf {k\alpha}}
\end{align}
In order to evaluate (\ref{energy}) in terms of potentials, we use the following normalization condition: 
\begin{align}\label{normalize}
\int\limits_{\mathcal{V}}e^{i\textbf{k}.\textbf{r}}e^{i{\bf{k}^\prime}.\textbf{r}}d^3r=V\delta_{\textbf{k},-{\bf{k}^\prime}}
\end{align}
Hence, we can write (\ref{energy}) in terms of potential as
\begin{align}
\mathcal{H}=&\frac{1}{2}\epsilon_o\int\limits_{\mathcal{V}}\left[|\textbf{E}|^2 +c^2|\textbf{B}|^2\right]dV \nonumber\\
=&\frac{1}{2}\epsilon_o\sum_{\substack{{\bf{k,k^\prime}}=-\infty \\ \alpha,\beta=1,2}}^{\infty}\int\limits_{\mathcal{V}}-\omega_k\omega_{k^\prime}\big({\bf{e}}_{\bf{k\alpha}}.{\bf{e}}_{\bf {k^\prime_\beta}}\big)\big(\tilde{b}_{\bf{k\alpha}}-\tilde{b}^*_{\bf{-k\alpha}}\big)\big(\tilde{b}_{\bf{k^\prime_\beta}}-\tilde{b}^*_{\bf{-k^\prime_\beta}}\big)e^{i({\bf{k}}+{\bf{k^\prime}}).\textbf{r}}dV+ \nonumber\\
&\frac{1}{2}\epsilon_o c^2 \sum_{\substack{{\bf{k,k^\prime}}=-\infty\\\alpha,\beta=1,2}}^{\infty}\int\limits_{\mathcal{V}}-\big\{{\bf{k}}\times{\bf{e}}_{\bf{k\alpha}}\big\}.\big\{{\bf{k^\prime}}\times{\bf{e}}_{\bf{k^\prime_\beta}}\big\}\big(\tilde{b}_{\bf{k\alpha}}+\tilde{b}^*_{\bf{-k\alpha}}\big)\big(\tilde{b}_{\bf{k^\prime_\beta}}+\tilde{b}^*_{\bf{-k^\prime_\beta}}\big)e^{i({\bf{k}}+{\bf{k^\prime}}).\textbf{r}}dV  \label{nasty}\\
=&\frac{1}{2}V\epsilon_o\omega_k^2
\sum_{\substack{{\bf{k}}=-\infty\\\alpha=1,2}}^{\infty}\big\{\big(\tilde{b}_{\bf{k\alpha}}-\tilde{b}^*_{-\bf{k\alpha}}\big)\big(\tilde{b}^*_{\bf{k\alpha}}-\tilde{b}_{-\bf{k\alpha}}\big)+\big(\tilde{b}_{\bf{k\alpha}}+\tilde{b}^*_{-\bf{k\alpha}}\big)\big(\tilde{b}^*_{\bf{k\alpha}}+\tilde{b}_{-\bf{k\alpha}}\big)\big\} \nonumber\\
=&\sum_{\substack{\textbf{k}=-\infty \\ \alpha=1,2}}^{\infty}V\epsilon_o\omega_k^2\left(\tilde{b}^*_{\bf{k\alpha}}\tilde{b}_{\bf{k\alpha}}+\tilde{b}_{\bf{k\alpha}}\tilde{b}^*_{\bf{k\alpha}}\right) \label{energypot1}\\
=&\sum_{\substack{\textbf{k}=-\infty \\ \alpha=1,2}}^{\infty}\frac{1}{2}\hbar\omega_k\left(\tilde{a}^*_{\bf{k\alpha}}\tilde{a}_{\bf{k\alpha}}+\tilde{a}_{\bf{k\alpha}}\tilde{a}^*_{\bf{k\alpha}}\right)
\label{energypot2}
\end{align}
where, we have used (\ref{normalize}) and ${\bf{k}}.{\bf{e}}_{\bf{k\alpha}}=0$ for simplifying (\ref{nasty}). We have
also redefined our variables as $\tilde{a}_{\bf{k\alpha}}=\sqrt{\frac{2V\epsilon_o\omega_k}{\hbar}}~\tilde{b}_{\bf{k\alpha}}$ 
in last step. 

Let us now define two variables analogous to position and momentum of Simple Harmonic Oscillator:
\begin{align}\label{QPvars}
\mathcal{Q}_{\bf{k\alpha}}&=\sqrt{\frac{\hbar}{2\omega_k}}\left(\tilde{a}^*_{\bf{k\alpha}}+\tilde{a}_{\bf{k\alpha}}\right) \\
\mathcal{P}_{\bf{k\alpha}}&=i\sqrt{\frac{\hbar\omega_k}{2}}\left(\tilde{a}^*_{\bf{k\alpha}}-\tilde{a}_{\bf{k\alpha}}\right)
\end{align} 
Note that, $\mathcal{Q},\mathcal{P}$ are real scalars depending upon direction of propagation and polarization. 
In terms of these variables the energy (\ref{energypot2}) can be  as:
\begin{align}
\mathcal{H}=&\sum_{\substack{\textbf{k}=-\infty \\ \alpha=1,2}}^{\infty}\frac{1}{2}
\left(\omega^2_k\mathcal{Q}^2_{\bf{k\alpha}}+\mathcal{P}^2_{\bf{k\alpha}}\right)
\end{align} 

So far our treatment has been totally classical. We now quantize this Hamiltonian taking 
$\mathcal{Q}_{\bf{k\alpha}}, \mathcal{P}_{\bf{k\alpha}}$ as the basic conjugate variables by following the standard 
quantization procedure of replacing each mode dependent position and momentum variables by corresponding operator. Note
that, we are not attempting to study Quantum Electrodynamics, but a quantum theory of the radiation modes. Since the
regime we are interested in is non-relativistic, this quantization is expected to provide a valid quantum theory.

The quantum mechanical Hamiltonian operator is given by:
\begin{align}\label{qmradhamil}
\mathcal{\hat{H}}=&\sum_{\substack{\textbf{k}=-\infty \\ \alpha=1,2}}^{\infty}\frac{1}{2}
\left(\omega^2_k\mathcal{\hat{Q}}^2_{\bf{k\alpha}}+\mathcal{\hat{P}}^2_{\bf{k\alpha}}\right)
\end{align} 
where, operators $\mathcal{\hat{Q}}_{\bf{k\alpha}},\mathcal{\hat{P}}_{\bf{k\alpha}}$ satisfy the following commutation relations
\begin{align}
\label{QPcommute}
&[\mathcal{\hat{Q}}_{\bf{k\alpha}},\mathcal{\hat{Q}}_{\bf{k\prime_\beta}}]=0 ~~~~
[\mathcal{\hat{P}}_{\bf{k\alpha}},\mathcal{\hat{P}}_{\bf{k\prime_\beta}}]=0 ~~~~
[\mathcal{\hat{Q}}_{\bf{k\alpha}},\mathcal{\hat{P}}_{\bf{k\prime_\beta}}]=i\hbar\delta_{kk\prime}\delta_{\alpha\beta} 
\end{align} 
These operators can further be written in terms of creation and annihilation operators 
using (\ref{QPvars}) as:
\begin{align}
\label{Qop}
\mathcal{\hat{Q}}_{\bf{k\alpha}}&=\sqrt{\frac{\hbar}{2\omega_k}}\left(\hat{a}^\dagger_{\bf{k\alpha}}+\hat{a}_{\bf{k\alpha}}\right) \\
\label{Pop}
\mathcal{\hat{P}}_{\bf{k\alpha}}&=i\sqrt{\frac{\hbar\omega_k}{2}}\left(\hat{a}^\dagger_{\bf{k\alpha}}-\hat{a}_{\bf{k\alpha}}\right)
\end{align} 
The canonical commutation relations (\ref{QPcommute}) imply that
\begin{align}
\label{aa*commute}
 &[\hat{a}_{\bf{k\alpha}},\hat{a}_{\bf{k\prime_\beta}}]=0 ~~~~
 [\hat{a}^\dagger_{\bf{k\alpha}},\hat{a}^\dagger_{\bf{k\prime_\beta}}]=0 ~~~~ [\hat{a}_{\bf{k\alpha}},\hat{a}^\dagger_{\bf{k\prime_\beta}}]=\delta_{kk\prime}\delta_{\alpha\beta} 
\end{align}
This allows us to rewrite the quantum Hamiltonian (\ref{qmradhamil}) in terms of
$\hat{a}_{\bf{k\alpha}},\hat{a}^\dagger_{\bf{k\alpha}}$ as:
 \begin{align}\label{qmradhamilaa*}
 \mathcal{\hat{H}}=&\sum_{\substack{\textbf{k}=-\infty \\ \alpha=1,2}}^{\infty}\left(\hat{a}^\dagger_{\bf{k\alpha}}\hat{a}_{\bf{k\alpha}}+\frac{1}{2}\right)\hbar\omega_k \nonumber\\
 =&\sum_{\substack{\textbf{k}=-\infty \\ \alpha=1,2}}^{\infty}\left(\hat{\mathcal{N}}_{\bf{k\alpha}}+\frac{1}{2}\right)\hbar\omega_k
 \end{align} 
where, $\hat{\mathcal{N}}_{\bf{k\alpha}}=\hat{a}^\dagger_{\bf{k\alpha}}\hat{a}_{\bf{k\alpha}}$ is photon number operator for corresponding radiation mode. 

Note that a naive quantization of (\ref{qmradhamilaa*}) will yield an infinite vacuum energy which is a classic problem 
encountered when one attempts to quantize radiation. Again, we are not interested in QED, so we will sidestep this problem by 
ignoring half factor in expression for Hamiltonian. The reason that it is sufficient for our purpose is that we are
ultimately interested in the problem of the leaky cavity in a laboratory setting. As we have seen in the damped
oscillator example, the time dependent decay of energy only depends  on the differences of energy levels matter. 
Since, we do not intend to calculate energies we will not discuss this problem further and just drop the half factor in
the subsequent calculations from the Hamiltonian (\ref{qmradhamilaa*}).

Moreover, in the laboratory, the measurable quantities are not the creation, annihilation operators 
$\hat{a}_{\bf{k\alpha}},\hat{a}^\dagger_{\bf{k\alpha}}$ but the electric and magnetic fields 
${{\bf \hat{E}}({\bf r},t)}, {{\bf \hat{B}}({\bf r},t)}$. Therefore for future reference we express the electric and
magnetic fields in terms of the creation and annihilation operators using (\ref{electric}) and (\ref{magnetic}) as
\begin{align}
\label{electricop}
{{\bf \hat{E}}({\bf r},t)}&=\sum_{\substack{\textbf{k}=-\infty \\ \alpha=1,2}}^{\infty}i\sqrt{\frac{\hbar\omega_k}{2V\epsilon_o}}\left [\hat{a}_{\bf{k\alpha}}e^{i\textbf{k}.\textbf{r}}-\hat{a}^\dagger_{\bf{-k\alpha}}e^{i\textbf{k}.\textbf{r}}\right]{\bf{e}}_{\bf {k\alpha}} \nonumber\\
&=\sum_{\substack{\textbf{k}=-\infty \\ \alpha=1,2}}^{\infty}i\sqrt{\frac{\hbar\omega_k}{2V\epsilon_o}}\left [\hat{a}_{\bf{k\alpha}}e^{i\textbf{k}.\textbf{r}}-\hat{a}^\dagger_{\bf{k\alpha}}e^{-i\textbf{k}.\textbf{r}}\right]{\bf{e}}_{\bf {k\alpha}} \\
\label{magneticop} 
{{\bf \hat{B}}({\bf r},t)}&=\sum_{\substack{\textbf{k}=-\infty \\ \alpha=1,2}}^{\infty}i\sqrt{\frac{\hbar}{2V\epsilon_o\omega_k}}~~\textbf{k}\times\left [\hat{a}_{\bf{k\alpha}}e^{i\textbf{k}.\textbf{r}}+\hat{a}^\dagger_{\bf{-k\alpha}}e^{i\textbf{k}.\textbf{r}}\right]{\bf{e}}_{\bf {k\alpha}} \nonumber\\
&=\sum_{\substack{\textbf{k}=-\infty \\ \alpha=1,2}}^{\infty}i\sqrt{\frac{\hbar}{2V\epsilon_o\omega_k}}~~\textbf{k}\times\left [\hat{a}_{\bf{k\alpha}}e^{i\textbf{k}.\textbf{r}}-\hat{a}^\dagger_{\bf{k\alpha}}e^{-i\textbf{k}.\textbf{r}}\right]{\bf{e}}_{\bf {k\alpha}}
\end{align}

We have, so far, obtained a quantum theory of electromagnetic radiation inside a perfectly conducting cavity. 
We now intend to introduce leakage term in the Hamiltonian (\ref{qmradhamilaa*}) and use the formalism developed in 
section (\ref{sec:Dissipat}) to understand how fields evolve with time in presence of leakage. 

\subsection{Radiation Decay}\label{subsec:raddecay}

In real world, it is not possible to construct boxes with perfectly conducting walls. Generally, the walls of a cavity
will have finite transmission and reflection coefficients such that the radiation leaks outside. 
That is what we plan to model in this section. This approach is novel and, to the best of our knowledge, has not been
reported in literature before.

Since we have written the Hamiltonian in a form analogous to the simple harmonic oscillator, we will attempt to model
the decay in a leaky cavity by introducing a complex term similar to the damped oscillator discussed previously. Explicitly
we modify the Hamiltonian (\ref{qmradhamilaa*}) as follows:
\begin{align}\label{leakyhamilt}
\mathcal{\hat{H}}_L=&\sum_{\substack{\textbf{k}=-\infty \\ \alpha=1,2}}^{\infty}\hat{\mathcal{N}}_{\bf{k\alpha}}
\hbar\omega_k - i\hbar\sum_{\substack{\textbf{k}=
-\infty \\ \alpha=1,2}}^{\infty}\gamma_{\bf{k\alpha}}\hat{\mathcal{N}}_{\bf{k\alpha}} \\
=&\mathcal{H} - i\hbar\sum_{\substack{\textbf{k}=-\infty \\ \alpha=1,2}}^{\infty}\gamma_{\bf{k\alpha}}\hat{\mathcal{N}}_{\bf{k\alpha}}
\end{align} 
where, $\gamma_{\bf{k\alpha}}$ is mode dependent damping factor with dimension of $[T]^{-1}$ and $\mathcal{H}$ is
undamped Hamiltonian. Note that $\gamma_{\bf{k\alpha}}$ can be different for different propagation and polarization directions. Note that, unlike
traditional treatments \cite{QOos} we do not consider the environment at all, we merely model the leakage by adding a
damping term by hand. The exact interaction of the environment with the cavity can be formulated by choosing specific
forms of $\gamma_{\bf{k\alpha}}$.
 
Since the creation operators for the different modes commute (see (\ref{aa*commute})) we can construct 
simultaneous eigenkets of operators $\hat{a}_{\bf{k\alpha}}$ as 
\begin{align}\label{leakycoherent}
\ket{\beta}=\prod_{\substack{\bigotimes \\ \textbf{k}=-\infty \\ \alpha=1,2}}^{k=\infty}\ket{\beta_{\bf{k\alpha}}} 
\end{align} 
where, $\bigotimes$ indicates direct product and $\ket{\beta_{\bf{k\alpha}}}$ is coherent state for corresponding mode given as
\begin{align}
\ket{\beta_{\bf{k\alpha}}}=e^{-\frac{|\beta_{\bf{k\alpha}}^2|}{2}}\sum\limits_{n=0}^\infty \frac{\beta_{\bf{k\alpha}}^n}{\sqrt{n!}}\ket{n}
\end{align} 
Note that now the problem has been completely mapped to the damped oscillator case which we had explicitly solved before.
The Hamiltonian (\ref{leakyhamilt}) is similar to the one we had written for the damped oscillator (\ref{dhohamilt}).
The coherent states (\ref{leakycoherent}) are simply generalization of the SHO coherent states (\ref{coherent}). We can
now follow the same steps as before: We determine the time dependent normalization factor from the time evolution of
the coherent states $\ket{\beta}$ by the non-hermitian Hamiltonian $\mathcal{\hat{H}}_L$. 
This will be used to construct the coherent states with time dependent normalization $\ket{\tilde{\beta},t}$. 
The expectation value of any observable will be calculated using these states.

The time evolution operator generated by Hamiltonian (\ref{leakyhamilt}) given by
$\hat{U}(t,0)=e^{-i\frac{\mathcal{\hat{H}}_L t}{\hbar}}$ evolves the states (\ref{leakycoherent}) as
\begin{align}
\ket{\beta,t}=U(t,0)\ket{\beta}=&\prod_{\substack{\bigotimes \\ \textbf{k}=-\infty \\ \alpha=1,2}}^{k=\infty}\ket{\beta_{\bf{k\alpha}},t} \\
=&\prod_{\substack{\bigotimes \\ \textbf{k}=-\infty \\ \alpha=1,2}}^{k=\infty}\left\{e^{-\frac{|\beta_{\bf{k\alpha}}^2|}{2}}\sum\limits_{n=0}^\infty \frac{\beta_{\bf{k\alpha}}^n}{\sqrt{n!}}e^{-\gamma_{\bf{k\alpha}} nt}e^{-i\frac{{E}_{n}t}{\hbar}}\ket{n}\right\}
\end{align}
The states which remain normalized for all times can be constructed from $\ket{\beta,t}$ using the time dependent
normalization
\begin{align}\label{leakyevol}
\ket{\tilde{\beta},t}=\prod_{\substack{\bigotimes \\ \textbf{k}=-\infty \\ \alpha=1,2}}^{k=\infty}\left\{e^{-\frac{|\beta_{\bf{k\alpha}}|^2}{2}e^{-2\gamma_{\bf{k\alpha}} t}}\sum\limits_{n=0}^\infty \frac{\beta_{\bf{k\alpha}}^n}{\sqrt{n!}}e^{-\gamma_{\bf{k\alpha}} nt}e^{-i\frac{{E}_{n}t}{\hbar}}\ket{n}\right\}
\end{align}

We can use these states to calculate the time dependence of any operator of interest. In particular 
the average photon number is given by
\begin{align}\label{leakyphoton}
\bra{\tilde{\beta},t}\hat{\mathcal{N}}\ket{\tilde{\beta},t}=&
\bra{\tilde{\beta},t}\sum_{\substack{\textbf{k}=-\infty \\ \alpha=1,2}}^{\infty}\hat{a}^\dagger_{\bf{k\alpha}}\hat{a}_{\bf{k\alpha}}\ket{\tilde{\beta},t} 
\nonumber\\
=\sum_{\substack{\textbf{k}=-\infty \\ \alpha=1,2}}^{\infty}&|\beta_{\bf{k\alpha}}|^2e^{{-2\gamma_{\bf{k\alpha}}t}}
\end{align}
As we can see that the photon number in the cavity decays with time. This expression also contains information about
selective decay of modes. To determine how much decay occurs in each mode we
can simply take $\gamma$ of the corresponding mode as non-zero number and all other $\gamma$'s as zero.

The quantities which can be measured are the electric and magnetic field. Their expectation values with respect to
$\ket{\tilde{\beta},t}$ can be evaluated using their expressions in (\ref{electricop}) and (\ref{magneticop}).  
\begin{align}
\label{electexp}
\bra{\tilde{\beta},t}{{\bf \hat{E}}({\bf r},t)}\ket{\tilde{\beta},t}=
&\sum_{\substack{\textbf{k}=-\infty \\ \alpha=1,2}}^{\infty}i\sqrt{\frac{\hbar\omega_k}{2V\epsilon_o}}
\left[\bra{\beta,t}\hat{a}_{\bf{k\alpha}}\ket{\beta,t}e^{i\textbf{k}.\textbf{r}}
-\bra{\beta,t}\hat{a}^\dagger_{\bf{k\alpha}}\ket{\beta,t}e^{-i\textbf{k}.\textbf{r}}\right]{\bf{e}}_{\bf {k\alpha}} 
\nonumber\\
=&\sum_{\substack{\textbf{k}=-\infty \\  \alpha=1,2}}^{\infty}-\sqrt{\frac{2\hbar\omega_k}{V\epsilon_o}}
~~\mbox{Im}\left(\beta_{\bf{k\alpha}}e^{i\textbf{k}.\textbf{r}}\right)e^{-\gamma_{\bf{k\alpha}}t}{\bf{e}}_{\bf{k\alpha}} 
\nonumber\\
=&\sum_{\substack{\textbf{k}=-\infty \\ \alpha=1,2}}^{\infty}{\bf E}({\bf {r}},0)_{\bf{k\alpha}}~e^{-\gamma_{\bf{k\alpha}}t} \\
\label{magexp}
\bra{\tilde{\beta},t}{{\bf \hat{B}}({\bf r},t)}\ket{\tilde{\beta},t}=
&\sum_{\substack{\textbf{k}=-\infty \\ \alpha=1,2}}^{\infty}i\sqrt{\frac{\hbar}{2V\epsilon_o\omega_k}}~~\textbf{k}\times
\left [\bra{\beta,t}\hat{a}_{\bf{k\alpha}}\ket{\beta,t}e^{i\textbf{k}.\textbf{r}}
-\bra{\beta,t}\hat{a}^\dagger_{\bf{k\alpha}}\ket{\beta,t}e^{-i\textbf{k}.\textbf{r}}\right]{\bf{e}}_{\bf {k\alpha}} 
\nonumber\\
=&\sum_{\substack{\textbf{k}=-\infty \\  \alpha=1,2}}^{\infty}-\sqrt{\frac{2\hbar}{V\epsilon_o\omega_k}}~~
\left[\mbox{Im}\left(\beta_{\bf{k\alpha}}e^{i\textbf{k}.\textbf{r}}\right)e^{-\gamma_{\bf{k\alpha}}t}\right]\textbf{k}\times{\bf{e}}_{\bf{k\alpha}} 
\nonumber\\
=&\sum_{\substack{\textbf{k}=-\infty \\ \alpha=1,2}}^{\infty}{\bf B}({\bf {r}},0)_{\bf{k\alpha}}~e^{-\gamma_{\bf{k\alpha}}t}
\end{align}
Hence, we see that the fields corresponding to mode with non-zero $\gamma$ decay with time. 
 
In Table 1 we plot the time decay of the electric field in a 1-D leaky cavity with 4 modes consisting of 
first, second, third and fourth harmonic with following dissipation factors:
\begin{align}
&\text{First Harmonic}:~~\gamma_1=0.0001; \nonumber\\
&\text{Second Harmonic}:~~\gamma_2=0.002; \nonumber\\
&\text{Third Harmonic}:~~\gamma_3=0.005;  \nonumber\\
&\text{Fourth Harmonic}:~~\gamma_4=0.01;  
\end{align}
\begin{table} 
\newcommand*{\addheight}[2][.5ex]{
  \raisebox{0pt}[\dimexpr\height+(#1)\relax]{#2}}
\centering
\begin{tabular}{|c|c|}
      \hline
      \addheight{\includegraphics[width=70mm]{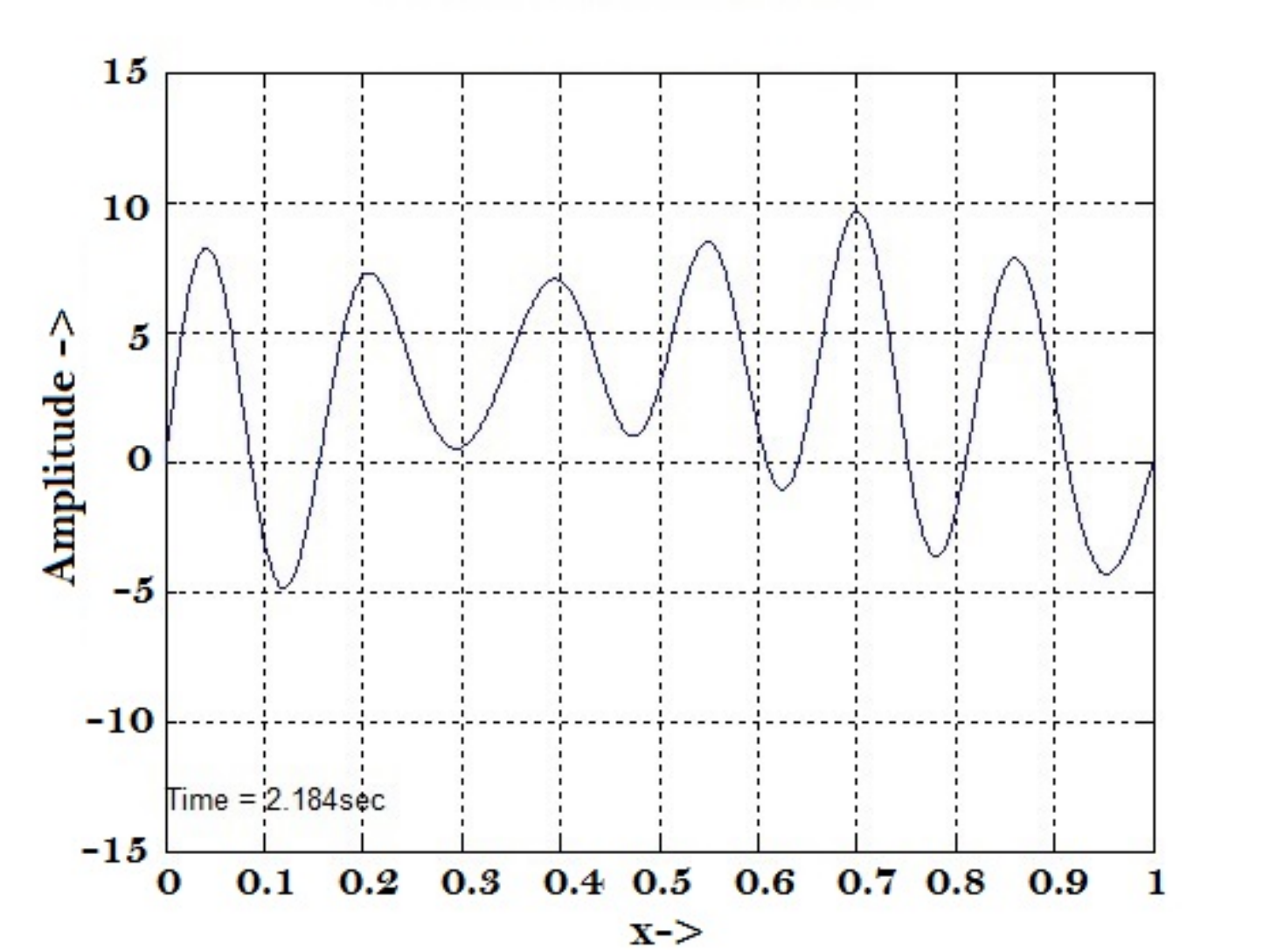}} &
      \addheight{\includegraphics[width=70mm]{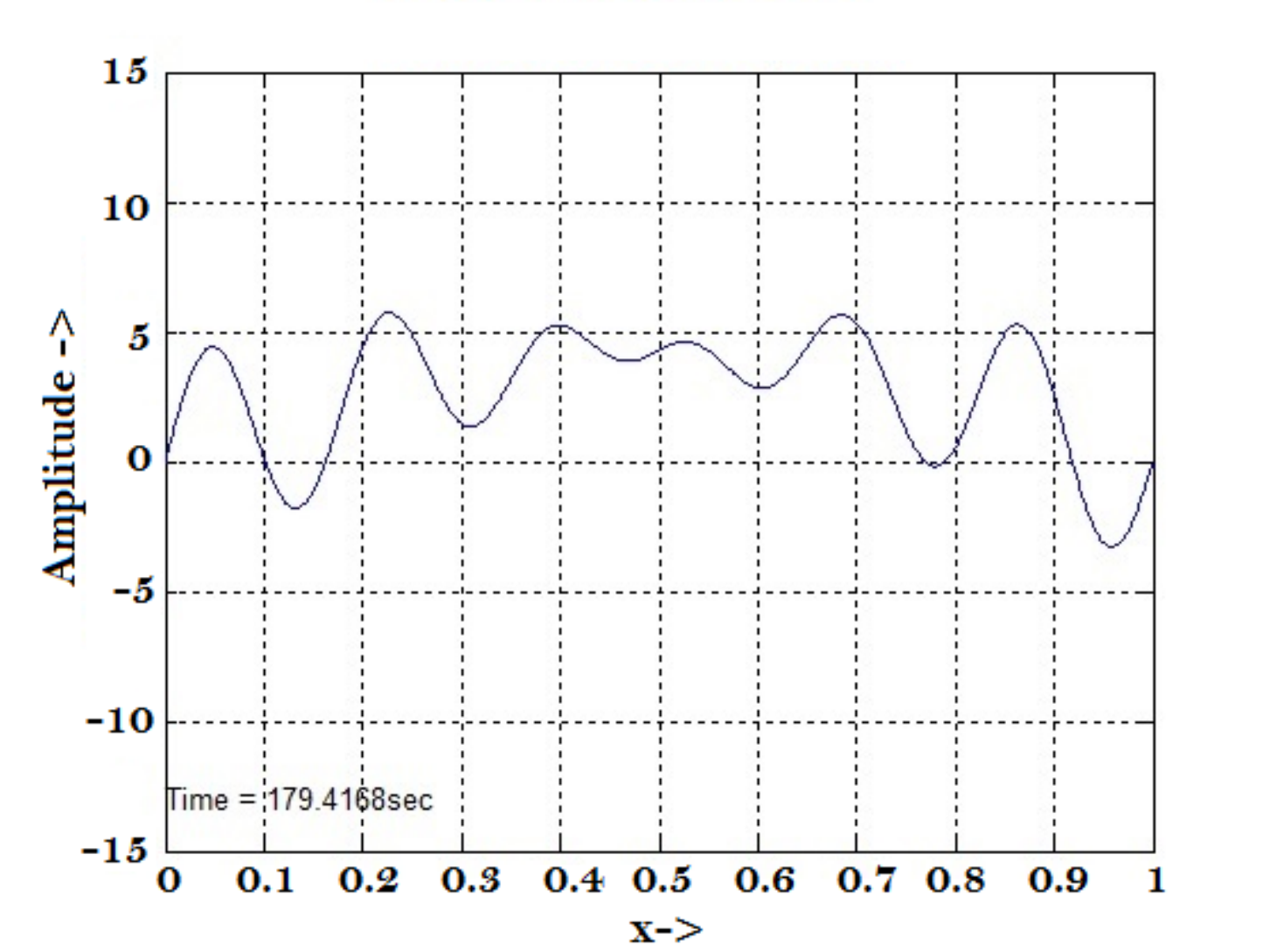}} \\
      \hline
      \small Electric field at t=2.18 sec& Electric field at t=179.42 sec \\
      \hline
      \addheight{\includegraphics[width=70mm]{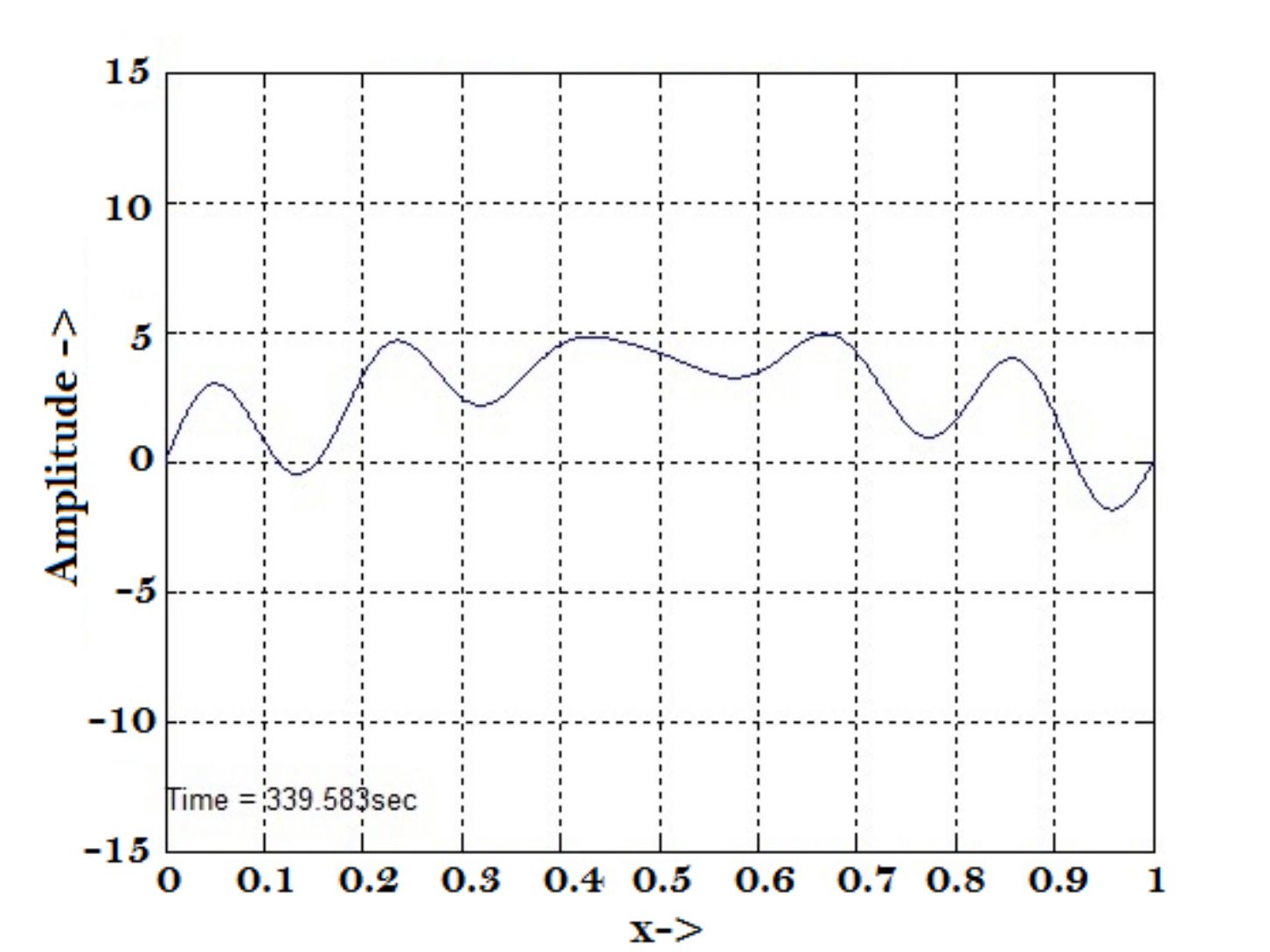}} &
      \addheight{\includegraphics[width=70mm]{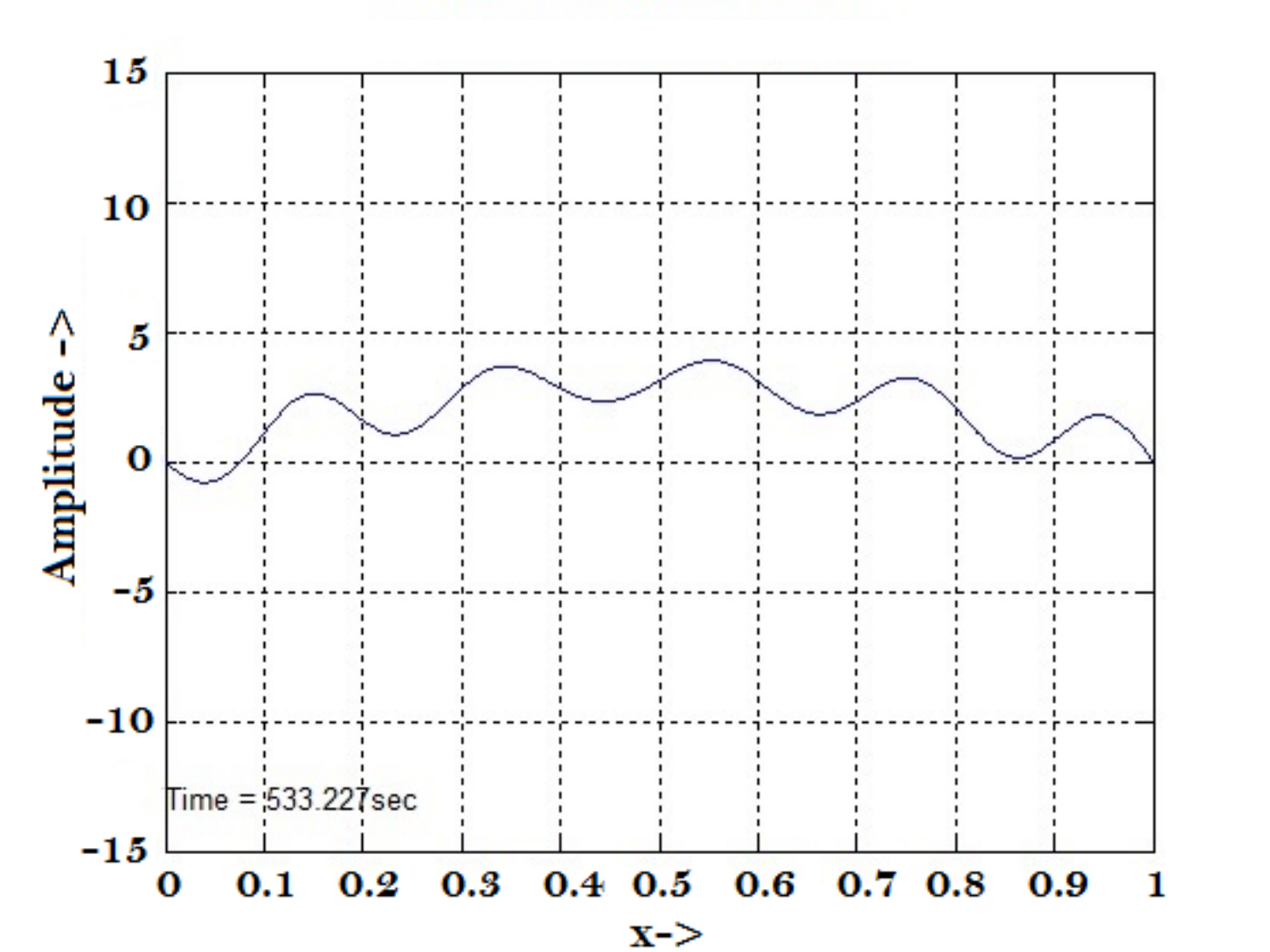}} \\
      \hline
      \small Electric field at t=339.58 sec& Electric field at t=533.23 sec \\
      \hline
\end{tabular}\label{tab:table1}
\caption{Time evolution of electric field of Electric Field}
\end{table}
As expected, the electric field tends towards zero as time progresses as more and more radiation leaks out.
We further observe that we can specialize our field expressions for the case of single mode cavity as studied in 
\cite{dutra} by taking $\gamma=\frac{\kappa}{2}$ where $\kappa$ depends upon cavity dimensions and boundary reflectivity. 
This gives us the following expression for fields
\begin{align}\label{singlemodeelect}
\bra{\tilde{\beta},t}{\bf \hat{E}}({\bf {r}},t)\ket{\tilde{\beta},t}
=&{\bf E}({\bf {r}},0)e^{-\frac{\kappa}{2} t} \\
\label{singlemodemagnetic}
\bra{\tilde{\beta},t}{\bf \hat{B}}({\bf {r}},t)\ket{\tilde{\beta},t}
=&{\bf B}({\bf {r}},0)e^{-\frac{\kappa}{2} t}
\end{align}
These results agree with the ones given in \cite{dutra}.

The formalism developed here is novel and much simpler than the ones existing in literature. Obviously the shape we have chosen
for our cavity (cuboid), makes the problem much simpler than the more realistic situations encountered in literature.
Generalizing this treatment to cavities of arbitrary shapes is a direction of future work. Another advantage of this formalism 
is that it can be easily generalized to study cavities whose walls allow only particular polarizations or particular
frequencies in specific directions to leak through. It also allows the analysis of leakage of multiple modes in one go. 
We think that this procedure can be a powerful analytical tool to study leaky cavities. 
\newpage

\section{Conclusion} \label{sec:conclude}

In this paper, we have made an improvement to the existing idea \cite{Graefe} of describing dissipative systems via non hermitian
Hamiltonians. The main difference from existing treatments is our use of coherent states with time dependent
normalization (\ref{normalcoh}). As we show in this paper, using these states greatly simplify the derivation of the
classical limit. Moreover, these states have a nice physical interpretation as they clearly show that in a dissipative
system, owing to the loss of energy to the environment, the probability of finding the system in the higher energy
states reduces with time. We demonstrate this procedure for two dissipative system, the standard damped harmonic
oscillator example and the problem of radiation in a leaky cavity. For the leaky cavity problem, we show that if the
cavity is cuboidal, the problem can be mapped on the damped oscillator system. Although the geometry of the chosen cavity
greatly simplified the problem, it is possible to apply the same formalism to cavities of any shape. That is a direction
of future work. The damped oscillator spectrum can also be mapped on to several dissipative systems. One of such systems
of particular interest is the mapping on to quasi-normal modes of a black hole \cite{shanki}.  Exploring our formalism in 
this and other general relativistic systems is one of the problems we are currently exploring.  
 
Another possible direction of future work using this formalism is addressing the problem of  quantum decoherence
\cite{decoh1}. Decoherence arises from the loss of information from the system to the surroundings consequently leading
to non-unitary time evolution of the system. One of the ways of modelling decoherence is through adding interference
terms linking the system and surrounding which ultimately drive the system to classicality \cite{decoh2}. Obviously non
hermitian Hamiltonians play a significant role in such processes. Whether the states we use in this paper help us
improve our understanding of decoherence is an interesting question for future research. 

\appendix
\section{Review of Simple Harmonic Oscillator}

For completeness,  we give a brief pedagogical review of quantum treatment of simple harmonic oscillator and coherent states. 
For detailed description reader can consult standard texts on quantum mechanics (see \cite{sakurai} for details).
The Hamiltonian operator for SHO is given as,
\begin{equation}\label{shohamilt}
\hat{H}=\frac{1}{2m}{\hat{\mathcal{P}}}^2+ \frac{\omega^2m}{2}{\hat{\mathcal{Q}}}^2
\end{equation}
where, $[\hat{\mathcal{Q}},\hat{\mathcal{P}}]=i\hbar$ is the canonical commutation relation (CCR). We define, annihilation operator $\hat{a}$ and 
creation operator ${\hat{a}}^\dagger$ as follows
\begin{align}\label{ladderop}
\hat{a}&=\sqrt{\frac{m\omega}{2\hbar}}\left[\hat{\mathcal{Q}}+\frac{i}{m\omega}{\hat{\mathcal{P}}}\right]\\
{\hat{a}}^\dagger&=\sqrt{\frac{m\omega}{2\hbar}}\left[\hat{\mathcal{Q}}-\frac{i}{m\omega}{\hat{\mathcal{P}}}\right]
\end{align}
where, $[\hat{a},\hat{a}^\dagger]=1$ due to CCR.

We further define number operator $\hat{N}=\hat{a}^\dagger\hat{a}$. We thus observe that Hamiltonian (\ref{shohamilt}) can be 
rewritten in terms of number operator as
\begin{equation}\label{shohamiltop}
\hat{H}=\left[\hat{N}+\frac{1}{2}\right]\hbar\omega
\end{equation}
Let us further consider eigenstate of number operator as $\ket{n}$ such that $N\ket{n}=n\ket{n}$ thus giving us,
\begin{equation}\label{shoenergy}
\hat{H}\ket{n}=\left[n+\frac{1}{2}\right]\hbar\omega=E_n\ket{n}
\end{equation}
The operators $\hat{a}$($\hat{a}\dagger$) act on energy state $\ket{n}$ by decreasing(increasing) the number $n$ by $1$.
\begin{align}\label{ladderalg}
\hat{a}\ket{n}&=\sqrt{n}\ket{n-1} \\
\hat{a}^\dagger\ket{n}&=\sqrt{n+1}\ket{n+1}
\end{align}
In order to establish connection with classical mechanics we can use minimum uncertainty states called as coherent
states. Coherent state is defined to be the eigenstates of operator $\hat{a}$. It can be expressed as a superposition of energy eigenstates as
\begin{align}\label{coherent}
\ket{\alpha}=e^{-\frac{|\alpha^2|}{2}}\sum\limits_{n=0}^\infty \frac{\alpha^n}{\sqrt{n!}}\ket{n}
\end{align}
Note that since, the operator $\hat{a}$ is non-hermitian the eigenvalue $\alpha$ will in general be complex. 


\end{document}